\pdfoutput=1
\documentclass[11pt,twoside,a4paper,cmspaper,final,collab]{cms-tdr}
\def\svnVersion{aa250f5}\def\svnDate{2023/01/20}\def\cmsCernNoTag{CERN-EP-2021-215}\def\cmsCernDate{\today}\def\cmsMessage{Published in Nature Physics as \href{http://dx.doi.org/10.1038/s41567-022-01838-y}{\doi{10.1038/s41567-022-01838-y}.}}
\begin{document}\cmsNoteHeader{BPH-21-004}

\newcommand{\TJPsi}{\JPsi\,\JPsi\,\JPsi}
\newcommand{\pp}{\ensuremath{\Pp\Pp}\xspace}
\newcommand{\mumu}{\ensuremath{\PGmp\PGmm}\xspace}
\newcommand{\massmumu}{\ensuremath{m_{\mumu}}\xspace}
\newcommand{\psio}{\ensuremath{\PGy_1}\xspace}
\newcommand{\psid}{\ensuremath{\PGy_2}\xspace}
\newcommand{\psit}{\ensuremath{\PGy_3}\xspace}
\newcommand{\xsec}{\sigma(\pp\to\TJPsi \,\PX)}
\newcommand{\ntjpsisig}{\ensuremath{N^{3\JPsi}_\text{sig}}\xspace}
\newcommand{\brjmm}{\ensuremath{\mathcal{B}_{\JPsi\to\mumu}}\xspace}
\newcommand{\xsecbrfull}{\ensuremath{\xsec = \ntjpsisig/(\epsilon\,\Lumi_\text{int}\,\mathcal{B}^{3}_{\JPsi\to\mumu})}\xspace}
\newcommand{\sigmaeffdps}{\ensuremath{\sigma_\text{eff,DPS}}\xspace}
\newcommand{\sigmaefftps}{\ensuremath{\sigma_\text{eff,TPS}}\xspace}
\newcommand{\expm}{\ensuremath{\,\text{(exp)}}\xspace}
\newcommand{\helaconia}{\textsc{HELAC-Onia}\xspace}

\hyphenation{final}
\newlength\cmsTabSkip\setlength{\cmsTabSkip}{1ex}
\providecommand{\cmsTable}[1]{\resizebox{\textwidth}{!}{#1}}
\ifthenelse{\boolean{cms@external}}{\providecommand{\cmsTableExtra}{}}{\providecommand{\cmsTableExtra}{\cmsTable}}
\ifthenelse{\boolean{cms@external}}{\providecommand{\cmsTableC}[1]{\resizebox{\textwidth}{!}{#1}}}{\providecommand{\cmsTableC}[1]{#1}}

\cmsNoteHeader{BPH-21-004}
\title{Observation of triple \texorpdfstring{\JPsi}{J/psi} meson production in proton-proton collisions}

\date{\today}

\abstract{
Protons consist of three valence quarks, two up-quarks and one down-quark, held together by gluons and a sea of quark-antiquark pairs. Collectively, quarks and gluons are referred to as partons. In a proton-proton collision, typically only one parton of each proton undergoes a hard scattering -- referred to as single-parton scattering -- leaving the remainder of each proton only slightly disturbed. Here, we report the study of double- and triple-parton scatterings through the simultaneous production of three \JPsi mesons, which consist of a charm quark-antiquark pair, in proton-proton collisions recorded with the CMS experiment at the Large Hadron Collider. We observed this process -- reconstructed through the decays of \JPsi mesons into pairs of oppositely charged muons -- with a statistical significance above five standard deviations. We measured the inclusive fiducial cross section to be $272^{+141}_{-104} \stat \pm 17 \syst\unit{fb}$, and compared it to theoretical expectations for triple-\JPsi meson production in single-, double- and triple-parton scattering scenarios. Assuming factorization of multiple hard-scattering probabilities in terms of single-parton scattering cross sections, double- and triple-parton scattering are the dominant contributions for the measured process.
}

\hypersetup{%
pdfauthor={CMS Collaboration},%
pdftitle={Observation of triple J/psi meson production in proton-proton collisions at sqrt(s) = 13 TeV},%
pdfsubject={CMS},%
pdfkeywords={CMS, QCD, triple parton scatterings, J/psi production}}

\maketitle 

High-energy particle accelerators are unique tools to study the structure of matter at the shortest distances. The most powerful accelerator today is the CERN Large Hadron Collider (LHC) that has so far collided beams of protons with energies of up to 6.5\TeV, resulting in center-of-mass energies of up to 13\TeV. Protons are used in energy-frontier colliders because they are relatively easy to accelerate and keep in a circular orbit to enable high collision rates. However, they are not fundamental particles and, in fact, have a complicated quantum mechanical structure. Protons consist of three quarks, two up-type and one down-type, and gluons, which hold together the three valence quarks, as well as of a ``sea'' of virtual quark-antiquark pairs, which are fundamental elements of the quantum vacuum. All these components are referred to as ``partons''. In the standard picture of proton-proton (\pp) collisions, typically only a few partons undergo a hard scattering with one another, with the remainders of each proton only slightly disturbed in the process. As the collision energy increases, the densities of gluons and sea quarks probed inside each proton grow rapidly. Thus, at high enough energies, more than one pair of partons can undergo a hard scattering in a single \pp collision, leading to the simultaneous and independent production of two or more particles with transverse momentum (\pt) and/or mass ($m$) above a few \GeV. Double-parton scatterings (DPS) were first observed at the CERN Intersecting Storage Rings some 35 years ago~\cite{AxialFieldSpectrometer:1986dfj,Sjostrand:1987su} and have been a subject of intense theoretical and experimental investigations ever since~\cite{Bartalini:2018qje}. Numerous DPS processes have been studied in measurements with many combinations of pairs of heavy and/or high-\pt produced particles~\cite{Bartalini:2018qje}. Triple-parton scatterings (TPS), where three hard parton interactions take place simultaneously, have only been proposed for study recently~\cite{dEnterria:2016ids} and have not been experimentally explored yet.

Studies of $n$-parton scattering (NPS) processes are important to elucidate the complicated inner structure of the proton and its evolution with energy~\cite{Diehl:2011yj,Blok:2011bu}. Many of these features, \eg, the parton density profile in the plane transverse to the colliding beams, as well as the various correlations (in position, momentum, flavor, color, spin, \etc) between individual partons, are very difficult to calculate theoretically, and can only be mapped out through experimental studies of NPS in different systems and for different numbers $n$ of simultaneous scatterings~\cite{Bartalini:2018qje}. Measurements of such processes not only allow for a deeper understanding of the proton structure, but are also of relevance at the LHC to predict backgrounds in rare standard model processes, and in searches for new physics, in final states where multiple heavy particles are produced~\cite{CMS:2019wch,Sirunyan:2020txn}. This work presents the experimental study of the simultaneous production of three massive particles originating in NPS, and the observation of triple-\JPsi meson production. The study of TPS via triple-\JPsi production provides input for further theoretical and experimental progress in understanding the NPS dynamics.

In the simplest approach, ignoring any correlations between the individual partons, the probability to produce $n$ high-\pt particles in a given \pp collision must be proportional to the $n^\text{th}$-product of probabilities to independently produce each of them. Thus, the probability to produce two or three high-\pt particles in DPS and TPS scales with the square and cube, respectively, of the corresponding single-parton scattering (SPS) probabilities~\cite{dEnterria:2017yhd}. The occurrence of DPS and TPS processes is therefore more likely for final states with large SPS cross sections, such as quarkonia states (\eg, \JPsi and $\Upsilon$ mesons), than for rarer heavier particles such as, \eg, electroweak (EW) bosons~\cite{Chapon:2020heu}. In the DPS case, the cross section to produce, \eg, two charmonium mesons \psio and \psid can be written as the product of the SPS cross sections for the production of each individual meson divided by an effective cross section to warrant the proper units of the final result,
\begin{linenomath}
\begin{equation}
\sigma^{\pp \to \psio\,\psid+\PX}_\text{DPS} = \left(\frac{\mathfrak{m}}{2}\right)\, \frac{\sigma^{\pp \to \psio+\PX}_\text{SPS} \, \sigma^{\pp\to \psid+\PX}_\text{SPS}}{\sigmaeffdps}.
\label{eq:pocketDPS}
\end{equation}
\end{linenomath} 
Here, $\mathfrak{m}$ is a combinatorial factor to avoid double counting, $\mathfrak{m}=1\,(2)$ if $\psio=\psid$ ($\psio\neq\psid$), and \sigmaeffdps is an effective cross section that, in a purely geometric approach, can be determined from the \pp transverse overlap~\cite{dEnterria:2017yhd}. A \textit{smaller} value of \sigmaeffdps, which is proportional to the average (squared) transverse separation of the partons participating in the two hard scatterings, implies \textit{larger} DPS yields.

For the proton form factors typically implemented in the \PYTHIA8~\cite{Sjostrand:2014zea} and \HERWIGpp~\cite{Seymour:2013qka} event generators commonly used in collider physics, values of $\sigmaeffdps\approx20\text{--}30\unit{mb}$ are expected. Such estimates are, however, about a factor of two larger than those experimentally derived via the ratio $\sigmaeffdps = (\sigma^{\pp \to \PX_1}_\text{SPS} \, \sigma^{\pp\to \PX_2}_\text{SPS})/\sigma^{\pp \to \PX_1\,\PX_2}_\text{DPS}$ for processes involving pairs of high-\pt jets and/or EW bosons, which are found to lie in the range $\sigmaeffdps\approx 10\text{--}20\unit{mb}$~\cite{ATLAS:2013aph,CMS:2013huw,ATLAS:2016rnd,CMS:2017han,Aaboud:2018tiq,CMS:2019jcb,CMS:2021lxi}. This discrepancy has been mostly explained as evidence of parton correlations in the collision not accounted for in the purely geometrical approaches~\cite{Blok:2017alw}. In addition, significantly lower $\sigmaeffdps\approx 3\text{--}10\unit{mb}$ values have been extracted from measurements of quarkonium pair production ($\JPsi\JPsi$~\cite{Aaij:2011yc,Abazov:2014qba,CMS:2014cmt,Aaboud:2016fzt,Aaij:2016bqq}, $\JPsi\Upsilon$~\cite{Abazov:2015fbl}, and $\Upsilon\Upsilon$~\cite{Khachatryan:2016ydm,Sirunyan:2020txn}) that have been interpreted as due to the different dominant species (mostly gluons for quarkonia, and quarks for EW bosons) in the parton distribution functions (PDFs) probed in the different scatterings~\cite{Bartalini:2018qje}, but can be also attributed in some cases to poorly controlled subtractions of SPS contributions~\cite{Chapon:2020heu}. 

\begin{figure*}[ht]
\centering
    \includegraphics[width=.99\textwidth]{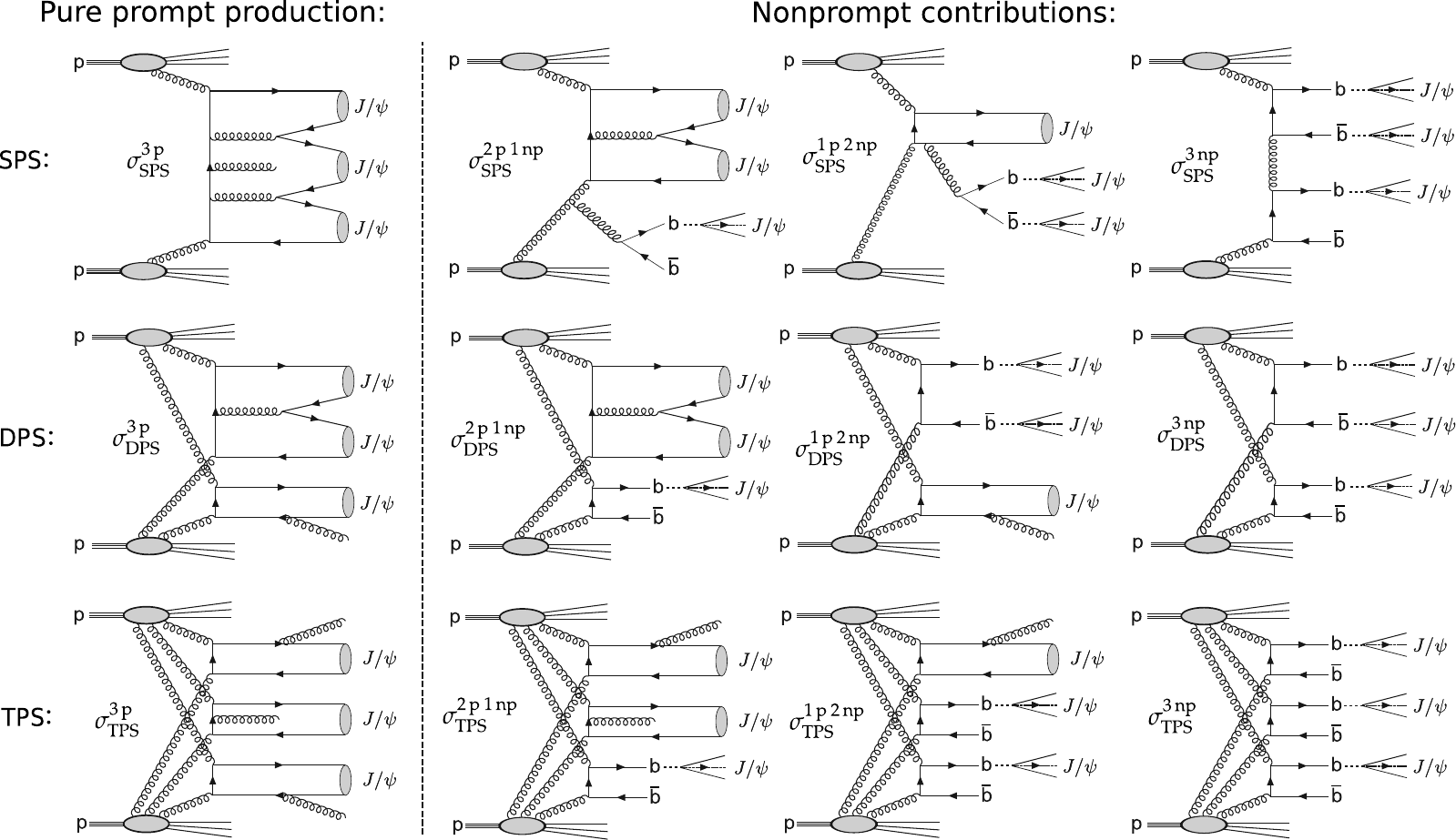}
    \vspace{0.2cm}
    \caption{Leading-order diagrams for inclusive triple-\JPsi production in \pp collisions. The SPS, DPS, and TPS processes are depicted in the top, middle, and bottom row, respectively. The leftmost diagrams show triple prompt-\JPsi processes. The remaining diagrams show (left to right) final states with increasing contributions of nonprompt \JPsi mesons from beauty hadron decays. Curly lines indicate gluons,
arrows (anti)charm quarks, and blobs the proton parton densities. The symbols $\sigma^{i\mathrm{p}j\mathrm{np}}_\mathrm{NPS}$ identify the number ($i$ and $j$) of prompt (p) and nonprompt (np) contributions to the cross section of each diagram.
    \label{fig:diags}}
\end{figure*}

The study of TPS via triple-\JPsi production can help solve all the issues mentioned above. The equivalent of Eq.~(\ref{eq:pocketDPS}) for the production of three charmonium mesons in a TPS process reads
\ifthenelse{\boolean{cms@external}}{
  \begin{linenomath}
    \begin{multline}
    \sigma^{\pp \to \psio\,\psid\,\psit+\PX}_\text{TPS} \\
    = \left(\frac{\mathfrak{m}}{3!}\right)\, \frac{\sigma^{\pp \to \psio+\PX}_\text{SPS} \, \sigma^{\pp\to \psid+\PX}_\text{SPS}\, \sigma^{\pp\to \psit+\PX}_\text{SPS}}{\sigmaefftps^2},
    \label{eq:pocketTPS}
    \end{multline}
    \end{linenomath}
}{
\begin{linenomath}
\begin{equation}
\sigma^{\pp \to \psio\,\psid\,\psit+\PX}_\text{TPS} = \left(\frac{\mathfrak{m}}{3!}\right)\, \frac{\sigma^{\pp \to \psio+\PX}_\text{SPS} \, \sigma^{\pp\to \psid+\PX}_\text{SPS}\, \sigma^{\pp\to \psit+\PX}_\text{SPS}}{\sigmaefftps^2},
\label{eq:pocketTPS}
\end{equation}
\end{linenomath} 
}
where $\mathfrak{m}=1,3$, or 6 (depending on whether all three, two, or none of the $\psi_i$ states are identical). In the absence of parton correlations, the effective cross section \sigmaefftps is closely related to its DPS counterpart via $\sigmaefftps = \kappa\,\sigmaeffdps$, with $\kappa$ of order unity. A value of $\kappa = 0.82\pm 0.11$ has been derived in~\cite{dEnterria:2016ids} for a variety of proton transverse parton profiles. A theoretical study of the production of three prompt-\JPsi mesons~\cite{Shao:2019qob}, based on the nonrelativistic quantum chromodynamics (NRQCD) approach at leading-order (LO) accuracy as implemented in the \helaconia code~\cite{Shao:2012iz,Shao:2015vga}, has demonstrated that the pure SPS contributions are negligible compared to those from DPS and TPS. Namely, the upper left diagram of Fig.~\ref{fig:diags} is irrelevant compared to the two other diagrams in the left column of the figure. The experimental measurement of $\pp\to\TJPsi\,\PX$ is thus a golden channel for the study of TPS and, in addition, provides an alternative extraction of \sigmaeffdps, thereby shedding light on the underlying dynamics of hard NPS. The production of \JPsi states can also proceed nonpromptly
through the decay of a beauty-quark (\PQb) hadron. Notwithstanding a small branching fraction, $\mathcal{B}_{\PQb\to\JPsi\,\PX}\approx 1\%$~\cite{ParticleDataGroup:2020ssz}, the cross section to produce \bbbar pairs is large at the LHC, $\sigma(\pp\to\bbbar\,\PX)\approx0.5\unit{mb}$~\cite{dEnterria:2016ids}. The contributions of such processes to inclusive triple-\JPsi production are schematically shown in Fig.~\ref{fig:diags} (diagrams to the right of the vertical dashed line).

This Letter reports the observation of the simultaneous production of three \JPsi mesons in \pp collisions. The analysis is based on a data sample collected at $\sqrt{s} = 13\TeV$ by the CMS experiment, corresponding to an integrated luminosity of 133\fbinv. The \JPsi mesons are reconstructed in their dimuon decay mode over a fiducial phase space in transverse momenta and (pseudo)rapidities ($\pt^{\mu,\JPsi},\,\abs{\eta^\PGm}$, and $\abs{y^{\JPsi}}$) defined to maximize the signal purity and the detector acceptance and efficiency. The analysis of the $6\PGm$ final state offers a very clean experimental signature for inclusive triple-\JPsi production, comprising prompt and nonprompt components. The Methods section provides more details on the experimental setup and event reconstruction.

\begin{figure*}[ht!]
\centering
    \includegraphics[width=.328\textwidth]{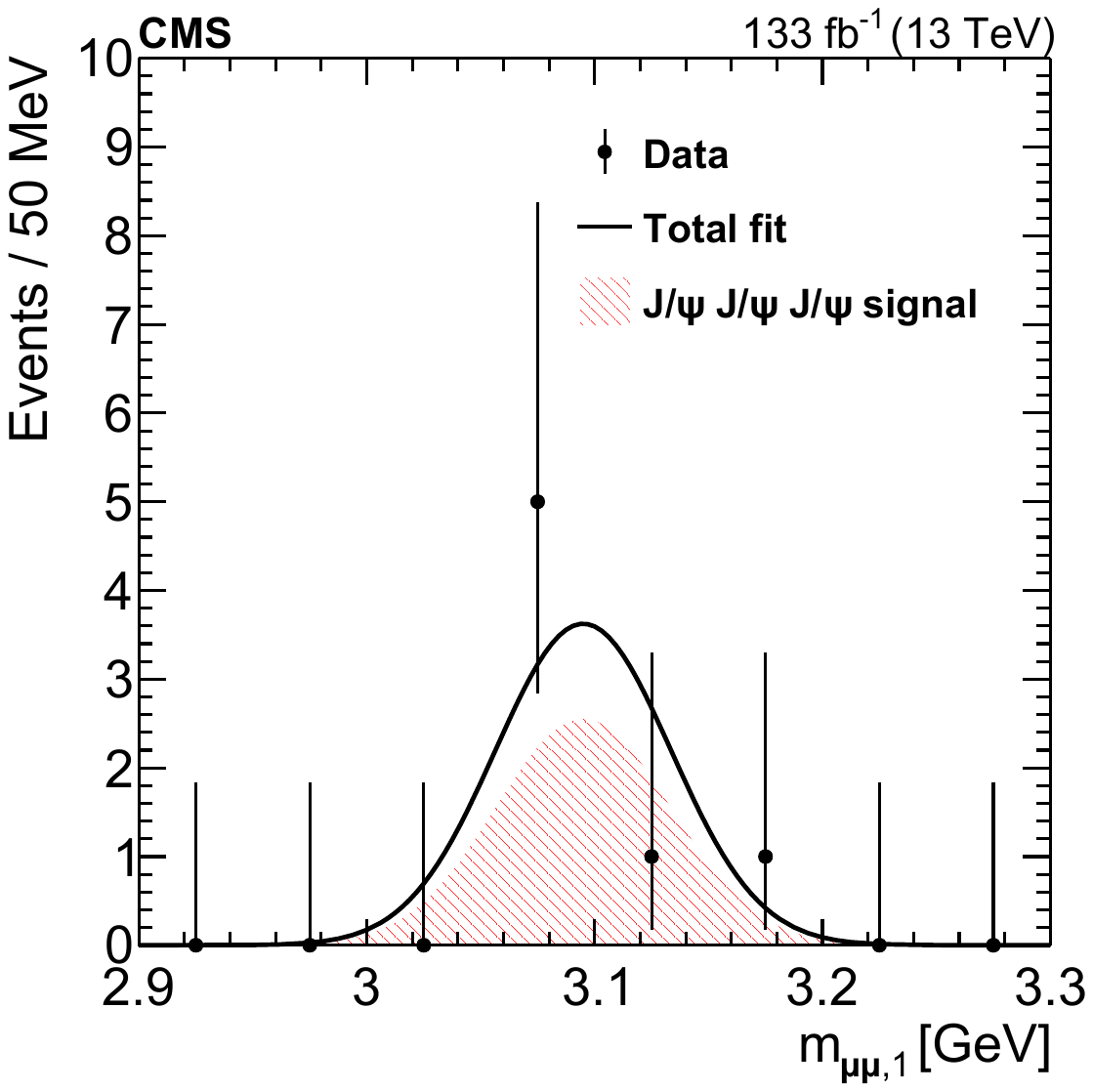}
    \includegraphics[width=.328\textwidth]{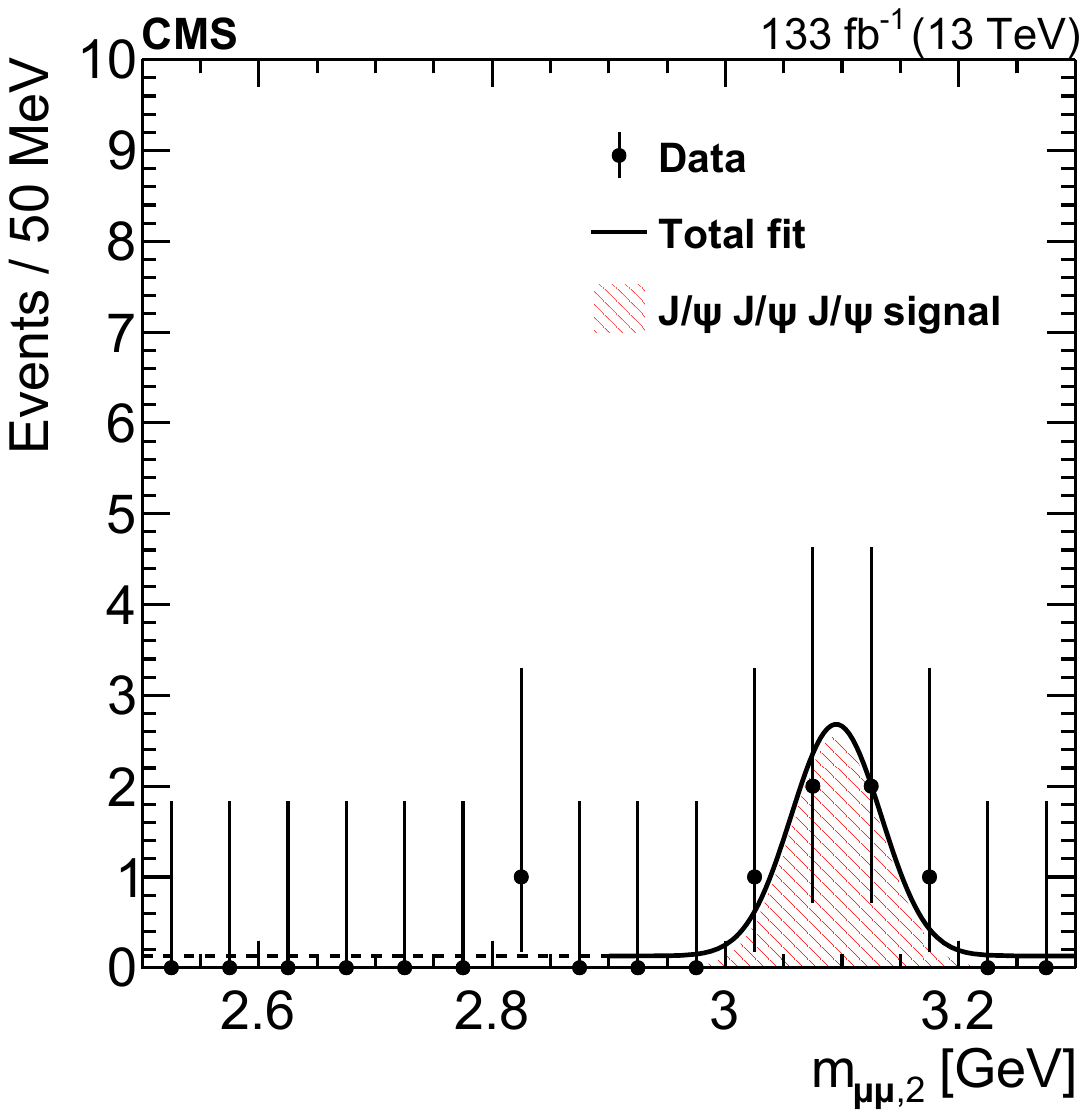}
    \includegraphics[width=.328\textwidth]{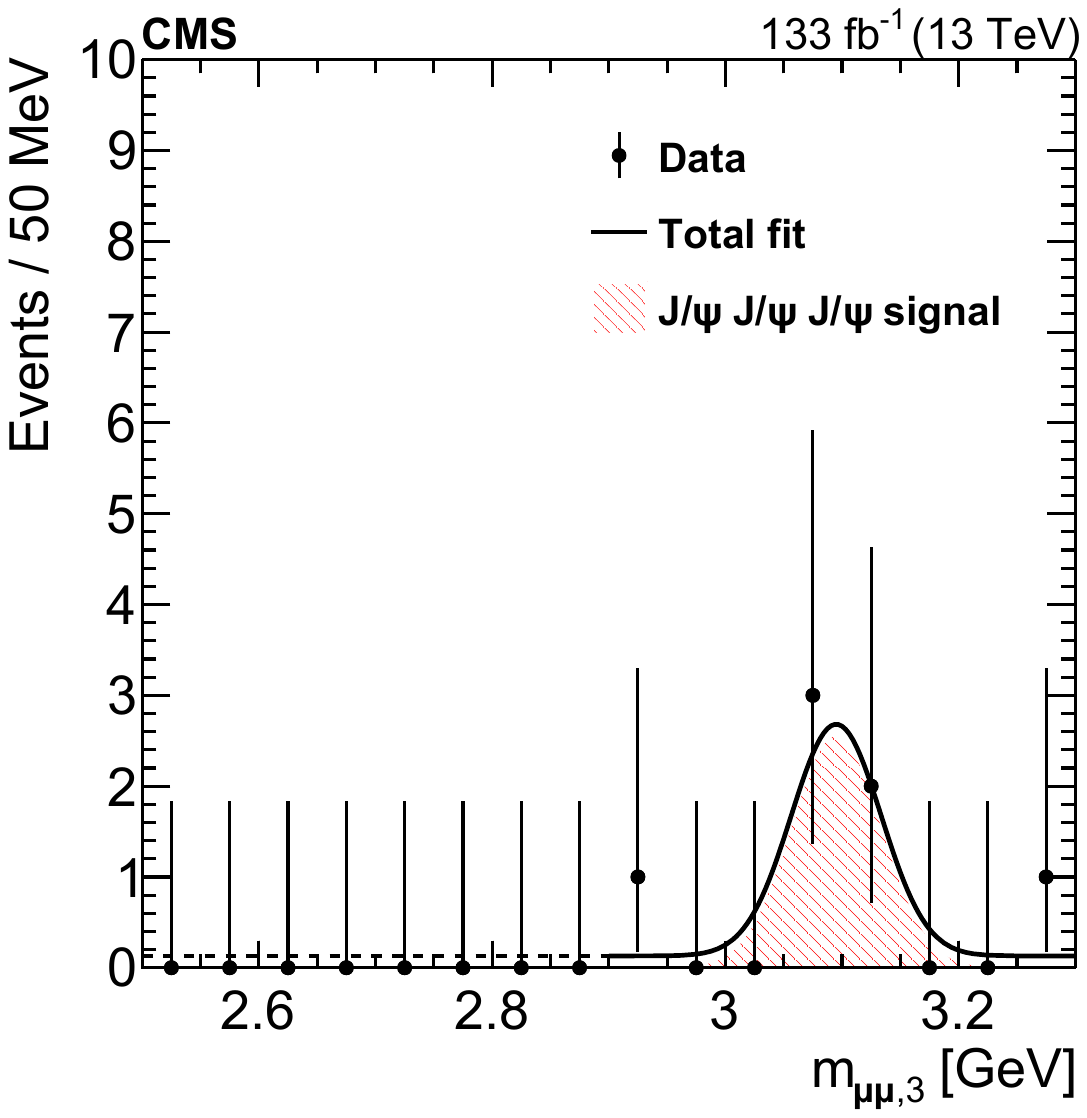}
    \caption{Invariant mass distributions for the three \mumu pairs in the selected events. The three distributions are ordered, left to right, by decreasing pair \pt. The data are represented by the points with the vertical bars showing the (Poisson) statistical uncertainties. The solid (dotted) curve shows the overall fit to the data (in the extended mass range, discussed in the main text), and the red shaded area the fitted signal yield. The horizontal bin width is indicated on the vertical axis legend.}
    \label{fig:fit}
\end{figure*}

The signal yield is extracted with a three-dimensional unbinned extended maximum likelihood fit of the \massmumu distributions of all \JPsi candidates in the event over the $2.9 < \massmumu < 3.3\GeV$ range. The expected \JPsi mass peaks are modeled with a Gaussian function with mean fixed to their nominal value ($m^{\JPsi} = 3.097\GeV$)~\cite{ParticleDataGroup:2020ssz} and the root-mean-square (RMS) width fixed to the resolution derived from the MC simulation ($\sigma_m \approx 30\MeV$). Given the very low number of events passing the selection, the mass mean and RMS width of the $\JPsi$ mesons cannot be left as free parameters in the fit. The dimuon background is described with an exponential function~\cite{CMS:2010nis,CMS:2011rxs,CMS:2014cmt,CMS:2015lbl,CMS:2017exb}. The fit has eight free parameters for the yields given by the combination of each of the three \JPsi candidates as being either signal or background. The extracted signal yield (red shaded areas in the \massmumu distributions of Fig.~\ref{fig:fit}) corresponds to $\ntjpsisig = 5.0^{+2.6}_{-1.9}$ triple-\JPsi events, with $1.0^{+1.4}_{-0.8}$ background events. The statistical significance of the signal is evaluated using various methods. From the likelihood ratio of two fits (background-only imposing $\ntjpsisig=0$, and the default signal-plus-background), with the standard asymptotic formula~\cite{Cowan:2010js} assuming that the conditions to apply Wilks' theorem~\cite{Wilks:1938dza} are satisfied, a significance of 6.7 standard deviations (std.\ dev.) is obtained. The significance derived assuming a Poisson counting experiment yields 5.8 std.\ dev., and it is found to be above 5.5 std.\ dev.\ by using MC pseudoexperiments.

To cross-check the size of the combinatorial background derived from the fit, two tests are carried out. First, the fit is repeated over the extended dimuon mass range [2.5--3.3]\GeV for the two subleading \JPsi mesons (dotted curves in Fig.~\ref{fig:fit}). This mass range corresponds to an asymmetric window of about $[-20\sigma_m,+7\sigma_m]$ around the \JPsi nominal mass that, as aforementioned, covers lower dimuon masses where the background, if any, should be larger. The obtained signal yield is fully consistent with the default result. A second test is performed whereby the opposite-sign (OS) requirement is removed to allow also for same-sign dimuon combinations ($\PGm^\pm\PGm^\pm$) for the two subleading pairs. After applying the rest of the selection criteria of the default analysis, no triplet events containing same-sign muon pairs are observed.

In order to estimate the average prompt and nonprompt contributions in the triple-\JPsi events, the proper decay-length of each \JPsi is  calculated as $L^{\JPsi}=(m^{\JPsi}/\pt^{\JPsi})\,L_{xy}^{\JPsi}$, where $L^{\JPsi}_{xy} =(\vec{r}\cdot\vec{\pt}^{\JPsi})/\abs{\pt^{\JPsi}}$ is the transverse distance between the \JPsi decay vertex and the PV ($\vec{r}$ is the vector from the PV to the \JPsi vertex). The experimental resolution on $L^\JPsi$ is ${\approx}20\mum$. Prompt \JPsi mesons are defined as those having $L^\JPsi<60\,\mum$, a choice that reduces the nonprompt contribution by more than one order in magnitude~\cite{CMS:2011rxs}. A signal-only subset of events is defined based on all \JPsi candidates found in a narrower invariant mass region, within $\pm3$ std.\ dev.\ around $m^{\JPsi}$, than that used in the default signal extraction. In this range, five triple-\JPsi events are found that can be classified as two events being consistent with the ``2 nonprompt $+$ 1 prompt \JPsi'' hypothesis, and one event each in the ``1 nonprompt $+$ 2 prompt \JPsi'', ``3 nonprompt \JPsi'', and ``3 prompt \JPsi'' categories. Such a classification is confirmed with a second method where prompt and nonprompt weights are extracted with the \textsc{sPlot} technique~\cite{Pivk:2004ty}, exploiting an unbinned maximum likelihood fit of the  $L^{\JPsi}$ distributions.

{\tolerance=800
The cross section for inclusive triple-\JPsi production measured in the fiducial region defined in Table~\ref{tab:fiducialregion} is obtained via \xsecbrfull, where \ntjpsisig is the number of signal events, $\Lumi_\text{int}$ the total integrated luminosity, \brjmm the dimuon branching fraction, and $\epsilon = \epsilon_\text{trig}\,\epsilon_\text{id} \, \epsilon_\text{reco}$ the total efficiency composed of trigger, identification, and reconstruction components. The \JPsi muon identification and reconstruction efficiencies are extracted with the tag-and-probe method using correction factors from the large inclusive $\JPsi\to\mumu$ data samples used in Ref.~\cite{Sirunyan:2018fpa}. Since they depend on the $\pt$ and $\eta$ of the muons, they propagate into the final cross section via two-dimensional maps, yielding $\epsilon_\text{id}\,\epsilon_\text{reco} = 0.78$. The trigger efficiency is found to be $\epsilon_\text{trig} = 0.84$ from a study of the MC samples.
\par}

\begin{table}[htpb!]
  \centering
  \topcaption{Fiducial phase space for the triple-\JPsi cross section measurement.}
  \begin{tabular}{lc}
  \hline
  \multirow{ 2}{*}{For all muons}
                    & $\pt>3.5\GeV$ for $\abs{\eta}<1.2$    \\
                    & $\pt>2.5\GeV$ for $1.2<\abs{\eta}<2.4$ \\ \hline
  \multirow{ 2}{*}{For all \JPsi mesons}
                    & $\pt>6\GeV$ and $\abs{y}<2.4$  \\
                    & $2.9<\massmumu<3.3\GeV$ \\ \hline
  \end{tabular}
\label{tab:fiducialregion}
\end{table}

The impact on the extracted cross section of the choice of functions used to reproduce the shapes of the signal and background dimuon invariant masses is studied. For the signal, the Gaussian distribution is changed to a Crystal-Ball function~\cite{Oreglia:1980cs} as well as to a Gaussian function with the RMS width left to vary in the fit. The background shape is changed from the default exponential to first- and zeroth-order polynomials. The relative differences in the cross sections obtained from the alternative modeling for signal and background are 0.8 and 3.4\%, respectively, and are assigned as corresponding systematic uncertainties. Uncertainties arising from the muon reconstruction and identification efficiencies are derived by allowing the tag-and-probe correction factors for each ($\pt,\eta$) bin to vary within their precision, and checking the effect on the extracted cross section. The maximal variation observed is $\pm1.0\%$. Varying the relative composition of double- and single-\JPsi meson production in the MC event sample used for the determination of the trigger efficiency leads to a 3.4\% propagated uncertainty. Uncertainties of 1.6\% and 3.0\% are added from the integrated luminosity measurement~\cite{CMS-LUM-17-003,CMS-PAS-LUM-17-004,CMS-PAS-LUM-18-002}, and from the simulated signal sample size, respectively. The uncertainty in the $\brjmm = (5.961 \pm 0.033)\%$ value~\cite{ParticleDataGroup:2020ssz} propagates into a $1.7\%$ uncertainty in the cross section. The total systematic uncertainty of the measured cross section is 6.2\%, obtained by adding all individual sources in quadrature (Extended Data Table~\ref{tab:systematics}). The measured cross section for triple-\JPsi production, within the fiducial region defined in Table~\ref{tab:fiducialregion}, is $\sigma(\pp\to\JPsi\JPsi\JPsi \,\PX)= 272^{+141}_{-104} \stat \pm 17 \syst\unit{fb}$. 

The total inclusive triple-\JPsi cross section is expected to correspond to the sum of the contributions from the SPS, DPS, and TPS processes schematically shown in Fig.~\ref{fig:diags}, each of which contains various combinations of prompt (p) and nonprompt (np) $\JPsi$ contributions,
\ifthenelse{\boolean{cms@external}}{
\begin{linenomath}
\begin{equation}
\begin{aligned}
\sigma^{3\JPsi}_\text{tot} = &
\,\,\sigma^{3\JPsi}_\text{SPS} + \sigma^{3\JPsi}_\text{DPS} + \sigma^{3\JPsi}_\text{TPS} \\
= &
\left(\sigma^{\mathrm{3p}}_\text{SPS} +
\sigma^{\mathrm{2p1np}}_\text{SPS} + \sigma^{\mathrm{1p2np}}_\text{SPS} +
\sigma^{\mathrm{3np}}_\text{SPS} \right)  \\ 
& +\left(\sigma^{\mathrm{3p}}_\text{DPS} +
\sigma^{\mathrm{2p1np}}_\text{DPS} + \sigma^{\mathrm{1p2np}}_\text{DPS} +
\sigma^{\mathrm{3np}}_\text{DPS} \right)  \\
& +\left(\sigma^{\mathrm{3p}}_\text{TPS} +
\sigma^{\mathrm{2p1np}}_\text{TPS} + \sigma^{\mathrm{1p2np}}_\text{TPS} +
\sigma^{\mathrm{3np}}_\text{TPS} \right).
\label{eq:sigma_3jpsi_tot}
\end{aligned}
\end{equation}
\end{linenomath}}
{
\begin{linenomath}
\begin{equation}
\begin{aligned}
\sigma^{3\JPsi}_\text{tot} = &
\,\,\sigma^{3\JPsi}_\text{SPS} + \sigma^{3\JPsi}_\text{DPS} + \sigma^{3\JPsi}_\text{TPS} \\
= &
\left(\sigma^{\mathrm{3p}}_\text{SPS} +
\sigma^{\mathrm{2p1np}}_\text{SPS} + \sigma^{\mathrm{1p2np}}_\text{SPS} +
\sigma^{\mathrm{3np}}_\text{SPS} \right)  \\ 
& + \left(\sigma^{\mathrm{3p}}_\text{DPS} +
\sigma^{\mathrm{2p1np}}_\text{DPS} + \sigma^{\mathrm{1p2np}}_\text{DPS} +
\sigma^{\mathrm{3np}}_\text{DPS} \right) +
\left(\sigma^{\mathrm{3p}}_\text{TPS} +
\sigma^{\mathrm{2p1np}}_\text{TPS} + \sigma^{\mathrm{1p2np}}_\text{TPS} +
\sigma^{\mathrm{3np}}_\text{TPS} \right).
\label{eq:sigma_3jpsi_tot}
\end{aligned}
\end{equation}
\end{linenomath}}
Under the simplest assumption of factorization of multiple hard-scattering probabilities in terms of SPS cross sections, the DPS and TPS contributions to triple-\JPsi production (last row of Eq.~(\ref{eq:sigma_3jpsi_tot})) can be written through Eqs.~(\ref{eq:pocketDPS}) and (\ref{eq:pocketTPS}) as a combination of products of single- and double-\JPsi SPS cross sections as
\ifthenelse{\boolean{cms@external}}{
  \begin{linenomath}
    \begin{gather}
    \begin{split}
    \sigma^{3\JPsi}_\mathrm{\mathrm{\textsc{dps}}} & =  \frac{\mathfrak{m}_1}{\sigmaeffdps}\\
    &\times\biggl[
    \sigma^\mathrm{2p}_\text{SPS}\sigma^\mathrm{1p}_\text{SPS}+
    \sigma^\mathrm{2p}_\text{SPS}\sigma^\mathrm{1np}_\text{SPS}+
    \sigma^\mathrm{1p}_\text{SPS}\sigma^\mathrm{1p1np}_\text{SPS} 
    \\
    &+
    \sigma^\mathrm{1p1np}_\text{SPS}\sigma^\mathrm{1np}_\text{SPS} +
    \sigma^\mathrm{1p}_\text{SPS}\sigma^\mathrm{2np}_\text{SPS} +
    \sigma^\mathrm{2np}_\text{SPS}\sigma^\mathrm{1np}_\text{SPS} 
    \biggr],
    \label{eq:sigma_3jpsi_dpstps_a}
    \end{split}
    \intertext{and}
    \begin{split}
    \sigma^{3\JPsi}_\mathrm{\textsc{tps}}  &= 
    \frac{\mathfrak{m}_3}{\sigmaefftps^2}\biggl[\left(\sigma^\mathrm{1p}_\text{SPS}\right)^3+\left(\sigma^\mathrm{1np}_\text{SPS}\right)^3\biggr]\\
    &+ \frac{\mathfrak{m}_2}{\sigmaefftps^2}\biggl[\left(\sigma^\mathrm{1p}_\text{SPS}\right)^2\sigma^\mathrm{1np}_\text{SPS}+\sigma^\mathrm{1p}_\text{SPS}\left(\sigma^\mathrm{1np}_\text{SPS}\right)^2\biggr],
    \label{eq:sigma_3jpsi_dpstps_b}
    \end{split}
  \end{gather}
\end{linenomath}
}{  
\begin{linenomath}
\begin{gather}
\begin{split}
\sigma^{3\JPsi}_\mathrm{\textsc{dps}}  = & \frac{\mathfrak{m}_1}{\sigmaeffdps}\,\biggl[
\sigma^\mathrm{2p}_\text{SPS}\sigma^\mathrm{1p}_\text{SPS}+
\sigma^\mathrm{2p}_\text{SPS}\sigma^\mathrm{1np}_\text{SPS}+
\sigma^\mathrm{1p}_\text{SPS}\sigma^\mathrm{1p1np}_\text{SPS} +
\sigma^\mathrm{1p1np}_\text{SPS}\sigma^\mathrm{1np}_\text{SPS} +
\sigma^\mathrm{1p}_\text{SPS}\sigma^\mathrm{2np}_\text{SPS} +
\sigma^\mathrm{2np}_\text{SPS}\sigma^\mathrm{1np}_\text{SPS} 
\biggr],
\label{eq:sigma_3jpsi_dpstps_a}
\end{split}
\intertext{and}
\begin{split}
\sigma^{3\JPsi}_\mathrm{\mathrm{\textsc{tps}}}  = & 
\frac{\mathfrak{m}_3}{\sigmaefftps^2}\biggl[\left(\sigma^\mathrm{1p}_\text{SPS}\right)^3+\left(\sigma^\mathrm{1np}_\text{SPS}\right)^3\biggr]+ \frac{\mathfrak{m}_2}{\sigmaefftps^2}\biggl[\left(\sigma^\mathrm{1p}_\text{SPS}\right)^2\sigma^\mathrm{1np}_\text{SPS}+\sigma^\mathrm{1p}_\text{SPS}\left(\sigma^\mathrm{1np}_\text{SPS}\right)^2\biggr],
\label{eq:sigma_3jpsi_dpstps_b}
\end{split}
\end{gather}
\end{linenomath}
}
with combinatorial prefactors $\mathfrak{m}_1 = 2/2=1, \mathfrak{m}_2 = 3/3! = 1/2$, and $\mathfrak{m}_3 = 1/3! = 1/6$. Therefore, from the eight individual SPS cross sections for single-, double-, and triple-\JPsi production, one can determine the total triple-\JPsi production cross section via Eqs.~(\ref{eq:sigma_3jpsi_tot}--\ref{eq:sigma_3jpsi_dpstps_b}). The values of the relevant SPS cross sections, each within the fiducial phase space defined in Table~\ref{tab:fiducialregion}, are computed as described next and listed in Table~\ref{tab:sigma_jpsi_theory}.

\begin{table*}[htpb!]
  \topcaption{Theoretical \JPsi, $\JPsi+\JPsi$, and $\JPsi+\JPsi+\JPsi$ cross sections. Predictions for \hbox{single-}, \hbox{double-}, and triple-\JPsi production cross sections in SPS processes, which pass the fiducial criteria listed in Table~\ref{tab:fiducialregion}, derived from the \helaconia (\textsc{ho})
  and \MGvATNLO (\textsc{mg5nlo}) matrix element calculators, complemented with the \PYTHIA8 (\textsc{py8}) parton shower, as described in the text. \label{tab:sigma_jpsi_theory}}
  \centering
  \renewcommand\arraystretch{1.5}
  \cmsTable{
  \begin{tabular}{cc|ccc|cccc} \hline
  \multicolumn{2}{c|}{SPS single-\JPsi production}  & \multicolumn{3}{c|}{SPS double-\JPsi production} & \multicolumn{4}{c}{SPS triple-\JPsi production} \\
  \textsc{ho(data)} & \textsc{mg5nlo+py8} & \textsc{ho(nlo*)} & \textsc{ho(lo)+py8} & \textsc{mg5nlo+py8} & \textsc{ho(lo)} & \textsc{ho(lo)+py8} & \textsc{ho(lo)+py8} & \textsc{mg5nlo+py8} \\\hline 
  $\sigma^\mathrm{1p}_\text{SPS}$ &  $\sigma^\mathrm{1np}_\text{SPS}$ &  $\sigma^\mathrm{2p}_\text{SPS}$ & $\sigma^\mathrm{1p1np}_\text{SPS}$ & $\sigma^\mathrm{2np}_\text{SPS}$ & $\sigma^\mathrm{3p}_\text{SPS}$ 
   & $\sigma^\mathrm{2p1np}_\text{SPS}$  &  $\sigma^\mathrm{1p2np}_\text{SPS}$ & $\sigma^\mathrm{3np}_\text{SPS}$ \\
  $570 \pm 57\unit{nb}$ & $600^{+130}_{-220}\unit{nb}$  & 
  $40^{+80}_{-26}\unit{pb}$  & $24^{+35}_{-16}\unit{fb}$  & $430^{+95}_{-130}\unit{pb}$ &
  $<5\unit{ab}$  & $5.2^{+9.6}_{-3.3}\unit{fb}$ & $14^{+17}_{-8}\unit{ab}$  & 
  $12\pm4\unit{fb}$ \\
  \hline
  \end{tabular}
  }
  \end{table*}

The values of the SPS single, double, and triple prompt-\JPsi cross sections ($\sigma^\mathrm{1p}_\text{SPS}$, $\sigma^\mathrm{2p}_\text{SPS}$, $\sigma^\mathrm{3p}_\text{SPS}$) are obtained with \helaconia at LO or at approximately next-to-leading-order (NLO*) accuracy~\cite{Lansberg:2013qka,Lansberg:2014swa}. For the single prompt-\JPsi prediction ($\sigma^\mathrm{1p}_\text{SPS}$), the theoretical calculations \textsc{ho(data)} in Table~\ref{tab:sigma_jpsi_theory} are normalized with a parameterization~\cite{Lansberg:2016deg} that reproduces the data measured in \pp collisions at 7\TeV~\cite{CMS:2015lbl,ATLAS:2015zdw}, including all feed-down contributions from decays of heavier charmonium resonances. For the double and triple prompt-\JPsi processes, the \helaconia calculations include only the $\Pgy$ feed-down because the $\PGcc$ decay contribution is only a few percent. All predictions for cross sections of nonprompt \JPsi meson production in beauty hadron decays ($\sigma^\mathrm{1np}_\text{SPS}$, $\sigma^\mathrm{2np}_\text{SPS}$, $\sigma^\mathrm{3np}_\text{SPS}$) have been obtained with \MGvATNLO (v.2.6.6)~\cite{MadGraph} matrix elements, scaled by a factor of 1.15 to account for next-to-next-to-leading-order \bbbar cross section corrections~\cite{dEnterria:2016ids}, combined with the \PYTHIA 8.244 generator for parton shower and decays (including all feed-down quarkonium contributions)~\cite{Lansberg:2014swa}. Mixed prompt plus nonprompt cross sections ($\sigma^\mathrm{XpYn}_\text{SPS}$) are obtained from $\JPsi+\bbbar$ events generated with \helaconia at LO in the color-singlet model interfaced with \PYTHIA 8.244 for the \PQb-quarks fragmentation into nonprompt \JPsi mesons. The uncertainties include the (dominant) theoretical scale dependence and the (subdominant) PDF uncertainties of the CT14NLO set~\cite{Dulat:2015mca}, except for the single prompt-\JPsi predictions that have a better precision because they are determined with an explicit fit of the NRQCD predictions to the LHC data~\cite{Lansberg:2016deg} and have an associated 10\% uncertainty of experimental origin. All these sources are treated as uncorrelated and the corresponding uncertainties are added in quadrature.

Using Eqs.~(\ref{eq:sigma_3jpsi_dpstps_a},\ref{eq:sigma_3jpsi_dpstps_b}) with the SPS cross sections listed in Table~\ref{tab:sigma_jpsi_theory}, and assuming that the effective DPS and TPS cross sections are related by $\sigmaefftps=(0.82\pm0.11)\,\sigmaeffdps$~\cite{dEnterria:2016ids} in a baseline approach that ignores parton correlations, one can extract the value of the effective DPS cross section that yields the experimentally measured $\sigma^{3\JPsi}_\text{tot}$ value. Following such a procedure, the value $\sigmaeffdps = 2.7^{+1.4}_{-1.0} \expm ^{+1.5}_{-1.0}\thy\unit{mb}$ is derived, where the first uncertainty is due to the experimental $\sigma^{3\JPsi}_\text{tot}$ precision and the second is due to the propagation of all sources of theoretical uncertainties in the ingredients of Eqs.~(\ref{eq:sigma_3jpsi_tot}--\ref{eq:sigma_3jpsi_dpstps_b}).

\begin{table*}[htpb!]
\topcaption{Expected contributions to triple-\JPsi production. Central values of the predictions for triple-\JPsi production cross sections (in \unit{fb}) and yields from SPS, DPS (for $\sigmaeffdps = 2.7\unit{mb}$), and TPS (for $\sigmaefftps=0.82\,\sigmaeffdps = 2.2\unit{mb}$) processes, and their total sum. The values are given in columns for combinations of $n$ prompt and $(3-n)$ nonprompt \JPsi mesons, the last column giving their corresponding sums. The DPS and TPS results are derived via Eqs.~(\ref{eq:sigma_3jpsi_tot}--\ref{eq:sigma_3jpsi_dpstps_b}) from the SPS cross sections listed in Table~\ref{tab:sigma_jpsi_theory} with $\sigmaeffdps$ chosen so that the sum of contributions yields a total cross section equal to the experimental value of $\sigma^{3\JPsi}_\text{tot}$. The expected yield, $N^{3\JPsi}_\mathrm{NPS}$, is given for an effective integrated luminosity of $\epsilon\,\Lumi_\text{int} = 87\fbinv$ for each contributing process. The last column lists the total cross sections and yields for SPS, DPS, TPS and their sums.
\label{tab:3_jpsi_components}}
\centering
\renewcommand\arraystretch{1.3}
\cmsTableExtra{
\begin{tabular}{lccccc} \hline
Process: & 3 prompt  & 2 prompt$+$1 nonprompt & 1 prompt$+$2 nonprompt  &  3 nonprompt  & Total \\\hline
$\sigma^{3\JPsi}_\text{SPS}$ (fb)  & ${<}0.005$ & $5.7$ & $0.014$ &  $12$ & $18$\\
$N^{3\JPsi}_\text{SPS}$ & 0.0 & 0.10 & 0.0 & 0.22 & 0.32 \\[\cmsTabSkip]
$\sigma^{3\JPsi}_\text{DPS}$ (fb)  & 8.4 & 8.9 & 90 & 95 & 202 \\
$N^{3\JPsi}_\text{DPS}$ & 0.15 & 0.16 & 1.65 & 1.75 & 3.7 \\[\cmsTabSkip]
$\sigma^{3\JPsi}_\text{TPS}$ (fb) & 6.1 & 19.4 &  20.4 & 7.2 & 53 \\ 
$N^{3\JPsi}_\text{TPS}$  & 0.11 & 0.36 & 0.38 & 0.13 & 1.0 \\[\cmsTabSkip]
$\sigma^{3\JPsi}_\text{tot}$ (fb)  & 15 & 34 & 110 & 114 & 272 \\
$N^{3\JPsi}_\text{tot}$ & 0.3 & 0.6 & 2.0 & 2.1 & 5.0 \\\hline
\end{tabular}
}
\end{table*}

The inclusive triple-\JPsi theoretical cross sections and yields for each individual process contributing to the total production are listed in Table~\ref{tab:3_jpsi_components}. The expected contributions from SPS, DPS, and TPS processes to the total triple-\JPsi cross section amount to about 6, 74, and 20\%, respectively. This confirms the conclusion of Ref.~\cite{Shao:2019qob} that triple-\JPsi production is a golden channel to study DPS and TPS, with minimal SPS contamination. The largest contributors to the triple-\JPsi cross section are $\sigma^\mathrm{3np}_\text{DPS}$ and $\sigma^\mathrm{1p2np}_\text{DPS}$, accounting for ${\approx}33\%$ each, $\sigma^\mathrm{2p1np}_\text{TPS}$ and $\sigma^\mathrm{1p2np}_\text{TPS}$ amounting to about 7\% each, and $\sigma^\mathrm{3np}_\text{SPS}$ representing about 4\% of the total production. In terms of prompt and nonprompt contributions, the theoretical expectation for the production of three promptly produced \JPsi mesons is $\approx$5\% of the total yield, whereas the percentage expected for three nonprompt \JPsi mesons is $\approx$45\%. The remaining half of the triple-\JPsi events are expected to be due to the combination of \JPsi mesons produced promptly and from beauty hadron decays. This result is consistent, within the large statistical uncertainties of the present data set, with the combination of prompt and nonprompt \JPsi mesons derived from the decay length of each dimuon candidate.

\begin{figure*}[htpb!]
\centering
    \includegraphics[width=.65\textwidth]{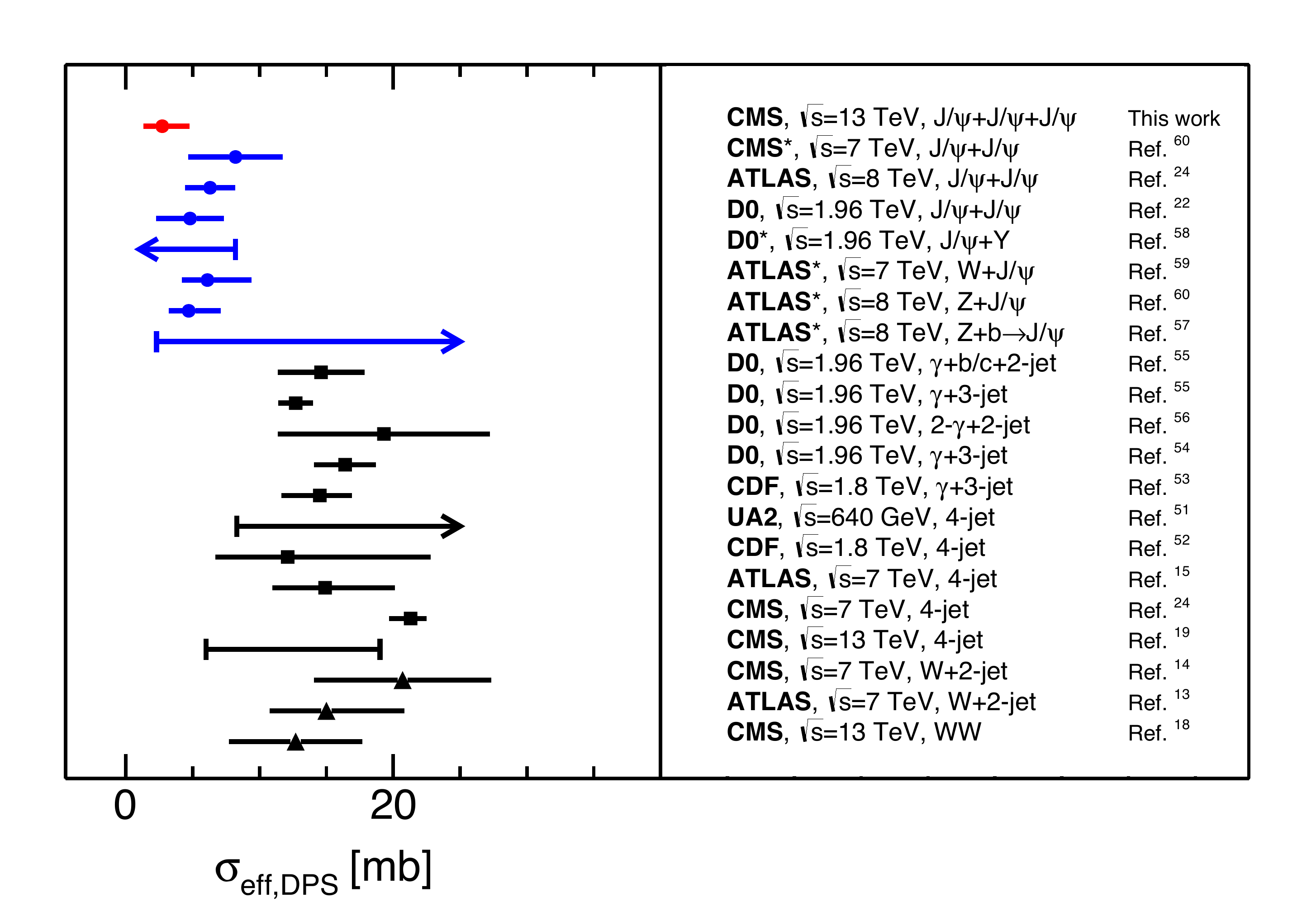}
    \caption{Comparison of $\sigmaeffdps$ parameters extracted in various processes. The result obtained here (upper red circle) is compared to those derived in midrapidity measurements of double-quarkonium and EW boson plus quarkonium production~\cite{Abazov:2014qba,Shao:2016wor,Aaij:2016bqq,Lansberg:2016muq,Aaboud:2016fzt,Lansberg:2017chq,Lansberg:2019adr} (blue circles), as well as in final states with jets~\cite{UA2:1991apc,CDF:1993sbj,Khachatryan:2015pea,CMS:2021lxi}, $\gamma+$\,jets~\cite{CDF:1997yfa,D0:2009apj,D0:2014owy,D0:2015rpo}, $\PW+$jets~\cite{ATLAS:2013aph,CMS:2013huw}, and same-sign \PW bosons~\cite{CMS:2019jcb} (black squares and triangles). The arrows indicate lower (or upper) limits at 95\% (68\%) confidence level. For the experimental results marked with a star, more recent theoretical reinterpretations based on more accurate calculations of the corresponding SPS cross section are plotted. The original experimental results can be found in Ref.~\cite{CMS:2014cmt} (CMS), Ref.~\cite{Abazov:2015fbl} (D0) and Refs.~\cite{ATLAS:2014yjd,ATLAS:2014ofp} (ATLAS). 
    \label{fig:sigma_eff}}
\end{figure*}

In Fig.~\ref{fig:sigma_eff}, the $\sigmaeffdps$ value extracted here (red circle) is compared to the world data on effective DPS cross sections derived from midrapidity measurements with quarkonium final states~\cite{ATLAS:2014ofp,Abazov:2015fbl,Abazov:2014qba,Aaboud:2016fzt,Aaij:2016bqq,Abazov:2014qba,Aaboud:2016fzt} (blue circles), as well as from processes with jets, photons, and/or $\PW$ bosons~\cite{UA2:1991apc,CDF:1993sbj,CDF:1997yfa,D0:2009apj,ATLAS:2013aph,CMS:2013huw,ATLAS:2016rnd,CMS:2017han,Aaboud:2018tiq,CMS:2019jcb,CMS:2021lxi,D0:2014owy,D0:2015rpo} (black squares and triangles). A few of the $\sigmaeffdps$ values plotted have been derived by more advanced phenomenological studies~\cite{Lansberg:2016muq,Shao:2016wor,Lansberg:2017chq,Lansberg:2019adr} of the experimental quarkonium data. The effective cross sections obtained from quarkonium measurements favor a smaller value of $\sigmaeffdps \approx 3\text{--}10\unit{mb}$ compared to the $\sigmaeffdps \approx 10\text{--}20\unit{mb}$ derived from harder or heavier final states. Such an apparent process-dependent $\sigmaeffdps$ value is suggestive of different parton transverse profiles, and/or correlations present, probed inside the proton at varying fractional momenta, given by $x = \sqrt{\smash[b]{{\pt}_{,\mathrm{V}}^2+m_{\mathrm{V}}^2}}\,\re^{\eta}/\sqrt{s}$, for $\mathrm{V}=\JPsi,\Upsilon,\PW,\PZ$. At midrapidity ($\abs{\eta}<2.5$), quarkonia are produced mostly in gluon-gluon scatterings carrying a fraction $x \approx 5\times10^{-4}$ of the proton momentum, whereas mostly quarks with $x\approx 10^{-2}$ participate in the production of EW bosons. The fact that LHCb measurements of double-quarkonia and quarkonia-plus-charm~\cite{LHCb:2012aiv,Aaij:2016bqq} at \textit{forward} rapidities ($\eta\approx2\text{--}4.5$), processes that originate in parton scatterings with asymmetric fractional momenta $x_{1} \approx 10^{-4}$ and $x_{2} \approx 10^{-2}$, lead to values of $\sigmaeffdps\approx 15\unit{mb}$, larger than those measured at midrapidity for similar final states, seems to confirm the dependence of the effective DPS cross section on the relevant parton species and $x$ fractions probed.
 
\bibliography{auto_generated}

\section*{Methods}

\subsection*{Experimental setup}

The central feature of the CMS apparatus is a superconducting solenoid of 6\unit{m} internal diameter, providing a magnetic field of 3.8\unit{T}. Within the solenoid volume are a silicon pixel and strip tracker, a lead tungstate crystal electromagnetic calorimeter, and a brass and scintillator hadron calorimeter, each composed of a barrel and two endcap sections. Forward calorimeters extend the $\eta$ coverage provided by the barrel and endcap detectors. Muons are measured over the range $\abs{\eta} < 2.4$ in gas-ionization detectors, embedded in the steel flux-return yoke outside the solenoid, made using three technologies: drift tubes, cathode strip chambers, and resistive plate chambers. A more detailed description of the CMS detector can be found in Ref.~\cite{Chatrchyan:2008zzk}.

Events of interest are selected using a two-tiered trigger system. The first trigger level, composed of custom hardware processors, uses information from the calorimeters and muon detectors to select events at a rate of around 100\unit{kHz} with a fixed latency of about 4\mus~\cite{Sirunyan:2020zal}. The second level (or high-level trigger, HLT) consists of a farm of processors running a version of the full event reconstruction software optimized for fast processing, and reduces the event rate to around 1\unit{kHz} before data storage~\cite{Khachatryan:2016bia}. The present analysis employs an HLT that requires three muons, each having $\pt>3.5\GeV$ for $\abs{\eta}<1.2$ (barrel) or $\pt>2.5\GeV$ for $1.2<\abs{\eta}<2.4$ (endcap). In addition, the event must have at least one pair of oppositely charged muons with invariant mass between 2.80 and 3.35\GeV that originate from a common vertex with a probability greater than 0.5\%, as determined by a Kalman vertex fit~\cite{Fruhwirth:1987fm}, thus suppressing random combinations of muons from unrelated sources. Monte Carlo (MC) simulations are used to determine the trigger efficiency. Simulated events are generated using \helaconia (v.2.6.6) with the NNPDF3.0 PDF set~\cite{NNPDF:2014otw}, and \PYTHIA 8.205 for the hadronization and decay with the CUETP8M1 underlying event tune~\cite{Khachatryan:2015pea}. Triple-\JPsi events are simulated by combining events from two different MC samples with varying relative fractions. The first MC sample contains three singly produced \JPsi mesons, whereas the second sample combines single- and double-\JPsi meson production, following Ref.~\cite{Shao:2019qob}. Since the \JPsi mesons originating in both processes have slightly different $\pt$ spectra, the variation of their relative fractions in the simulation allows to effectively scan the triple-muon trigger efficiency in the more uncertain region close to the trigger $\pt$ thresholds. The generated events are then processed through a detailed \GEANTfour simulation~\cite{Agostinelli:2002hh} of the CMS detector response.

\subsection*{Event reconstruction and selection}

Muons are reconstructed by combining information from the silicon tracker and the muon system~\cite{Sirunyan:2018fpa}. The matching between tracks reconstructed in each of the subsystems proceeds either outside-in, starting from a track in the muon system, or inside-out, starting from a track provided by the silicon tracker. Matching muons to tracks measured in the silicon tracker leads to a relative \pt resolution of 1\% in the barrel and 3\% in the endcaps for muons with \pt up to 100\GeV~\cite{Sirunyan:2018fpa}. The candidate vertex with the largest value of summed physics-object $\pt^2$ in the event is taken to be the primary \pp interaction vertex (PV)~\cite{CMS:2017yfk}. Only charged particles that are either directly participating in the PV determination, or are closest to the PV along the $z$ direction and not used in any other PV identification (namely, they are consistent with secondary decays from the closest PV), are used in the analysis. Simulation studies show that this procedure efficiently suppresses any potential contamination from charged particles produced in any other pp collision taking place in the same bunch crossing (pileup).

The offline data analysis starts by selecting events with six or more reconstructed muons, each passing the \pt and $\eta$ criteria implemented at the HLT. The muons are combined into OS pairs, and are considered for further study if their invariant mass is $2.9 < \massmumu < 3.3\GeV$ (corresponding to about $\pm7$ times the \JPsi mass resolution discussed below) and originate from a common vertex with a probability greater than 0.5\%, thereby reducing the random pairing of muons originating from other background sources in the same event, as discussed below. All selected muon pairs are further required to share the same PV (including the possibility that they originate from secondary vertices associated to a common PV) to eliminate accidental combinations of muons from different \pp collisions in the same or neighboring bunch crossings. The analysis thereby includes prompt-\JPsi mesons coming directly from the \pp interaction (or from feed-down decays of promptly produced and decayed resonances), and nonprompt ones coming from the decays of beauty hadrons. In order to ensure high purity and reconstruction efficiency, the \JPsi candidates are required to have $\pt>6\GeV$ and $\abs{y}<2.4$. 

After applying the selection criteria discussed above, six triple-\JPsi events are observed. 
No alternative pairings among the OS muons in these six events are found to satisfy the analysis requirements. In all events, the \JPsi meson with highest \pt (``leading'' \JPsi) is found to correspond to the muon pair that passed the online trigger selection. In inclusive \JPsi measurements~\cite{CMS:2010nis,CMS:2011rxs,CMS:2014cmt,CMS:2015lbl,CMS:2017exb}, a continuum background is present in the $m_{\mumu}$ distribution due to random combinations of OS muons originating from semileptonic beauty- and charm-quark hadron decays, $\PQb\to\PGm+\PX$ and $\PQc\to\PGm+\PX$, and Drell--Yan events, which pass the trigger and data analysis criteria. To monitor the sideband background population, the mass windows corresponding to the two subleading \JPsi mesons are extended well below the \JPsi mass region. The corresponding dimuon invariant mass distributions are shown in Fig.~\ref{fig:fit} ordered (left to right) by decreasing \pt of the \mumu system. Only one additional background event is found in the extended mass region indicated by the dotted curves, confirming that the combinatorial continuum is suppressed by the requirement of having three reconstructed \JPsi candidates in the same event.

\subsection*{Event kinematic properties}

Detailed information of the kinematic properties of the \JPsi candidates passing the triple-\JPsi selection criteria are shown in Extended Data Table~\ref{tab:jpsimesoninfo}. The kinematic distributions of all six \JPsi triplets do not show any local peak, which could be indicative of, \eg, a resonance decaying into three \JPsi mesons, but are distributed featureless (with large point-to-point statistical uncertainties) over triple-\JPsi mass, \pt, and $\eta$.

All sources of systematic uncertainty in the triple-\JPsi cross section measurement are listed in Extended Data Table~\ref{tab:systematics}.

\begin{table*}[htpb!]
\centering
  \topcaption{Dimuon invariant mass, proper decay-length, transverse momentum, rapidity, and azimuthal angle of each of the three \JPsi candidates measured in the six triple-\JPsi events passing our selection criteria.  (The last \JPsi candidate in the last row, with invariant mass $m^{\JPsi,3} = 2.94\GeV$, is likely a background event).}
  \renewcommand\arraystretch{1.0}
  \begin{tabular}{lcccccc}
  \hline \\
  Event & $m^{\JPsi,1}$\,(\GeVns) & $m^{\JPsi,2}$\,(\GeVns) & $m^{\JPsi,3}$\,(\GeVns)  & $L^{\JPsi,1}\,\unit{(mm)}$ & $L^{\JPsi,2}\,\unit{(mm)}$ & $L^{\JPsi,3}\,\unit{(mm)}$ \\ 
  \hline \\
  1     & 3.08                & 3.10                & 3.07                              & $\phantom{-}1.77$         & 0.24                      & $-0.01$               \\
  2     & 3.15                & 3.06                & 3.09                              & $\phantom{-}0.05$         & 0.36                      & $\phantom{-}0.02$     \\ 
  3     & 3.10                & 3.14                & 3.11                              & $-0.04$                   & 0.03                      & $\phantom{-}0.05$     \\ 
  4     & 3.07                & 3.03                & 3.09                              & $\phantom{-}0.48$         & 0.81                      & $\phantom{-}0.82$     \\ 
  5     & 3.12                & 3.14                & 3.14                              & $-0.25$                   & 0.13                      & $-0.02$               \\ 
  6     & 3.06                & 3.17                & 2.94                              & $\phantom{-}0.11$         & 0.38                      & $\phantom{-}0.61$     \\ 
  \hline \\ 
  Event & $\pt^{\JPsi,1}$\,(\GeVns) & $\pt^{\JPsi,2}$\,(\GeVns)   & $\pt^{\JPsi,3}$\,(\GeVns) & $y^{\JPsi,1}$ & $y^{\JPsi,2}$ & $y^{\JPsi,3}$  \\ 
  \hline \\ 
  1     & 17.64                 & 17.50                   & 8.68                  & $-2.25$             & $-0.39$           & $\phantom{-}1.53$ \\ 
  2     & 91.50                 & 54.04                   & 11.81                 & $\phantom{-}1.99$   & $\phantom{-}0.81$ & $-0.71$           \\ 
  3     & 11.29                 & 10.29                   & 6.98                  & $-0.50$             & $-0.37$           & $-1.64$           \\
  4     & 15.46                 & 10.61                   & 7.84                  & $-0.83$             & $-2.24$           & $-1.78$           \\ 
  5     & 8.67                  & 7.71                    & 6.75                  & $\phantom{-}2.03$   & $-1.14$           & $-1.87$           \\ 
  6     & 60.70                 & 19.09                   & 17.03                 & $\phantom{-}1.59$   & $\phantom{-}2.29$ & $\phantom{-}1.58$ \\ 
  \hline \\ 
  Event  & $\phi^{\JPsi,1}$ & $\phi^{\JPsi,2}$ & $\phi^{\JPsi,3}$ \\
  \cline{1-4} \\ 
  1      & $-1.98$            & $\phantom{-}2.06$   & $-1.56$     \\ 
  2      & $\phantom{-}2.60$  & $-2.14$             & $-0.38$      \\ 
  3      & $-0.87$            & $-1.50$             & $\phantom{-}0.66$  \\ 
  4      & $\phantom{-}1.00$  & $-2.07$             & $-1.77$     \\ 
  5      & $-1.77$            & $\phantom{-}1.99$   & $\phantom{-}2.91$  \\ 
  6      & $-1.98$            & $-1.74$             & $-2.17$     \\ 
  \cline{1-4} 
  \end{tabular}
\label{tab:jpsimesoninfo}
\end{table*}

All sources of systematic uncertainty in the triple-\JPsi cross section measurement are listed in Extended Data Table~\ref{tab:systematics}.

\begin{table}[htpb!]
  \centering
  \topcaption{Relative contributions to the systematic uncertainty of the $\xsec$ measurement. The last row gives the sum in quadrature of all components.}
  \cmsTableC{
  \begin{tabular}{lc}
  \hline
  Source                            &  Relative uncertainty   \\ \hline
  \JPsi meson signal shape          &  0.8\%   \\
  Dimuon continuum background shape &  3.4\%   \\
  Muon reconstruction and identification   &  1.0\%   \\
  Trigger efficiency                &  3.4\%   \\
  MC sample size                    &  3.0\%   \\
  Integrated luminosity             &  1.6\%   \\
  Dimuon branching fraction         &  1.7\%   \\\hline
  Total                             &  6.2\%   \\ \hline
  \end{tabular}
  }
\label{tab:systematics}
\end{table}

\section*{Data availability}

Tabulated results are provided in the HEPData record for this analysis~\cite{hepdata}. Release and preservation of data used by the CMS Collaboration as the basis for publications is guided by the CMS policy as stated in \href{https://cms-docdb.cern.ch/cgi-bin/PublicDocDB/RetrieveFile?docid=6032&filename=CMSDataPolicyV1.2.pdf&version=2}{CMS data preservation, re-use and open access policy}.

\section*{Code availability}

The CMS core software is publically available at \href{https://github.com/cms-sw/cmssw}{https://github.com/cms-sw/cmssw}.

\begin{acknowledgments}
  We congratulate our colleagues in the CERN accelerator departments for the excellent performance of the LHC and thank the technical and administrative staffs at CERN and at other CMS institutes for their contributions to the success of the CMS effort. In addition, we gratefully acknowledge the computing centers and personnel of the Worldwide LHC Computing Grid and other centers for delivering so effectively the computing infrastructure essential to our analyses. Finally, we acknowledge the enduring support for the construction and operation of the LHC, the CMS detector, and the supporting computing infrastructure provided by the following funding agencies: BMBWF and FWF (Austria); FNRS and FWO (Belgium); CNPq, CAPES, FAPERJ, FAPERGS, and FAPESP (Brazil); MES and BNSF (Bulgaria); CERN; CAS, MoST, and NSFC (China); MINCIENCIAS (Colombia); MSES and CSF (Croatia); RIF (Cyprus); SENESCYT (Ecuador); MoER, ERC PUT and ERDF (Estonia); Academy of Finland, MEC, and HIP (Finland); CEA and CNRS/IN2P3 (France); BMBF, DFG, and HGF (Germany); GSRI (Greece); NKFIA (Hungary); DAE and DST (India); IPM (Iran); SFI (Ireland); INFN (Italy); MSIP and NRF (Republic of Korea); MES (Latvia); LAS (Lithuania); MOE and UM (Malaysia); BUAP, CINVESTAV, CONACYT, LNS, SEP, and UASLP-FAI (Mexico); MOS (Montenegro); MBIE (New Zealand); PAEC (Pakistan); MSHE and NSC (Poland); FCT (Portugal); JINR (Dubna); MON, RosAtom, RAS, RFBR, and NRC KI (Russia); MESTD (Serbia); SEIDI, CPAN, PCTI, and FEDER (Spain); MOSTR (Sri Lanka); Swiss Funding Agencies (Switzerland); MST (Taipei); ThEPCenter, IPST, STAR, and NSTDA (Thailand); TUBITAK and TAEK (Turkey); NASU (Ukraine); STFC (United Kingdom); DOE and NSF (USA).
\end{acknowledgments}
\cleardoublepage \appendix\section{The CMS Collaboration \label{app:collab}}\begin{sloppypar}\hyphenpenalty=5000\widowpenalty=500\clubpenalty=5000\cmsinstitute{Yerevan~Physics~Institute, Yerevan, Armenia}
A.~Tumasyan
\cmsinstitute{Institut~f\"{u}r~Hochenergiephysik, Vienna, Austria}
W.~Adam\cmsorcid{0000-0001-9099-4341}, J.W.~Andrejkovic, T.~Bergauer\cmsorcid{0000-0002-5786-0293}, S.~Chatterjee\cmsorcid{0000-0003-2660-0349}, K.~Damanakis, M.~Dragicevic\cmsorcid{0000-0003-1967-6783}, A.~Escalante~Del~Valle\cmsorcid{0000-0002-9702-6359}, R.~Fr\"{u}hwirth\cmsAuthorMark{1}, M.~Jeitler\cmsAuthorMark{1}\cmsorcid{0000-0002-5141-9560}, N.~Krammer, L.~Lechner\cmsorcid{0000-0002-3065-1141}, D.~Liko, I.~Mikulec, P.~Paulitsch, F.M.~Pitters, J.~Schieck\cmsAuthorMark{1}\cmsorcid{0000-0002-1058-8093}, R.~Sch\"{o}fbeck\cmsorcid{0000-0002-2332-8784}, D.~Schwarz, S.~Templ\cmsorcid{0000-0003-3137-5692}, W.~Waltenberger\cmsorcid{0000-0002-6215-7228}, C.-E.~Wulz\cmsAuthorMark{1}\cmsorcid{0000-0001-9226-5812}
% ...
\vskip\cmsinstskip
% ...
1:~Also~at TU~Wien, Wien, Austria\\
% ...
\end{sloppypar}
\end{document}